\documentclass[runningheads]{llncs}
\usepackage[T1]{fontenc}
\usepackage{amssymb}
\usepackage{amsmath}
\usepackage{graphicx}
\usepackage{bbm}
\usepackage{booktabs}
\usepackage{multirow}
\usepackage{multicol}
\usepackage[flushleft]{threeparttable}
\usepackage{caption}
\usepackage{subfig}
\usepackage{stackengine}
\usepackage{makecell}
\usepackage{hyperref}
\usepackage{color}
\usepackage[misc]{ifsym}
\usepackage[square,sort,comma,numbers]{natbib}  
\begin{document}
%
\title{Mesh-based 3D Motion Tracking in Cardiac MRI using Deep Learning}
%
\titlerunning{Mesh-based 3D motion tracking}
%
\author{Qingjie Meng\inst{1}(\Letter) \and
Wenjia Bai\inst{1,2,3} \and
Tianrui Liu\inst{4} \and \\
Declan P O'Regan\inst{5} \and
Daniel Rueckert\inst{1,6}}
%
\authorrunning{Q. Meng et al.}
%
%
\institute{\footnotesize BioMedIA, Department of Computing, Imperial College London, UK \and
Data Science Institute, Imperial College London, UK \and
Department of Brain Sciences, Imperial College London, UK \and
National University of Defense Technology, China \and
MRC London Institute of Medical Sciences, Imperial College London, UK \and
Klinikum rechts der Isar, Technical University Munich, Germany\\
\email{q.meng16@imperial.ac.uk}}
\maketitle              
\begin{abstract}
3D motion estimation from cine cardiac magnetic resonance (CMR) images is important for the assessment of cardiac function and diagnosis of cardiovascular diseases. Most of the previous methods focus on estimating pixel-/voxel-wise motion fields in the full image space, which ignore the fact that motion estimation is mainly relevant and useful within the object of interest, \emph{e.g.}, the heart. In this work, we model the heart as a 3D geometric mesh and propose a novel deep learning-based method that can estimate 3D motion of the heart mesh from 2D short- and long-axis CMR images. By developing a differentiable mesh-to-image rasterizer, the method is able to leverage the anatomical shape information from 2D multi-view CMR images for 3D motion estimation. The differentiability of the rasterizer enables us to train the method end-to-end. One advantage of the proposed method is that by tracking the motion of each vertex, it is able to keep the vertex correspondence of 3D meshes between time frames, which is important for quantitative assessment of the cardiac function on the mesh. We evaluate the proposed method on CMR images acquired from the UK Biobank study. Experimental results show that the proposed method quantitatively and qualitatively outperforms both conventional and learning-based cardiac motion tracking methods.

\end{abstract}
\section{Introduction}
Estimating left ventricular (LV) myocardial motion is important for the detection of LV dysfunction and the diagnosis of myocardial diseases~\cite{Claus2015,Puyol2019}.
Most of recent cardiac motion tracking approaches utilize cine cardiac magnetic resonance (CMR) images to estimate a dense motion field which represents pixel-/voxel-wise deformation across the entire image, \emph{e.g.},~\cite{Qin2018,Bello2019,Puyol2019,Qin2020,Bai2020,Yu2020,Ye2021,Loecher2021}.
However, it remains challenging to use this type of motion estimation for certain clinical applications where the heart and its motion needs to be represented on a 3D geometric mesh with known vertex correspondence across time frames, \emph{e.g.}, pathological cardiac remodeling~\cite{Mansi2011} and motion-based characterization of LV phenotypes~\cite{Marvao2015}. 
Transforming image space pixel-/voxel-wise deformation to mesh space vertex deformation has several limitations. In specific, a pixel-wise motion field only considers the motion of the heart within a single view plane and does not provide complete 3D motion information. Using post-processing steps to convert pixel-/voxel-wise motion fields to 3D mesh displacements can compromise motion estimation accuracy due to interpolation operation and the lack of through-plane motion.

In this work, we propose an end-to-end trainable network for estimating the 3D motion of the myocardial mesh from cine CMR images. 
Our method integrates short-axis (SAX) and long-axis (LAX) view images, which enables estimating both in-plane and through-plane mesh motion. In the proposed method, the intensity information of the input multi-view images are utilized to directly estimate a 3D mesh displacement. 
The estimated mesh displacement explicitly models the 3D motion of each vertex from the end-diastolic (ED) frame to the $t$-th frame, and thus is able to maintain vertex correspondence between time frames. 
A differentiable mesh-to-image rasterizer is introduced during training to generate 2D soft segmentations from the mesh, which allows leveraging 2D multi-view anatomical shape information for 3D mesh displacement estimation. 
During inference, our model generates a sequence of meshes, which shows the myocardial motion across the cardiac cycle.
Extracting different anatomical view planes from the meshes can further generate 2D myocardial segmentations.

The main contributions of this paper are summarized as follows: (1) We address the problem of 3D cardiac motion tracking on the geometric mesh. The proposed method learns vertex-wise displacements of the 3D myocardial mesh directly from 2D SAX and LAX cine CMR images via an end-to-end trainable network. (2) We introduce a differentiable mesh-to-image rasterizer to leverage 2D anatomical shape information from different anatomical views for 3D motion estimation. (3) The proposed method is able to perform joint cardiac motion estimation and segmentation across the cardiac cycle.

\textit{\textbf{Related work.}}
Many cardiac motion estimation methods use conventional and deep learning-based image registration to estimate pixel-/voxel-wise motion fields~\cite{Rueckert1999,Vercauteren2007,McLeod2011,Craene2012,Qin2018,ZhengQ2019,Ta2020,Yu2020}. 
For example, a free form deformation (FFD) method that proposed by Rueckert et al.~\cite{Rueckert1999} for non-rigid image registration has been used for cardiac motion estimation in many recent works, \emph{e.g.},~\cite{Shi2012,Tobon2013,Bello2019,Puyol2019,Bai2020}. Vercauteren et al.~\cite{Vercauteren2007} introduced a non-parametric diffeomorphic image registration method which has been used for cardiac motion tracking~\cite{Qin2020}. In recent years, deep learning-based image registration methods have also been applied to cardiac motion estimation: Qin et al.~\cite{Qin2018} proposed a joint deep learning network for simultaneous cardiac segmentation and motion estimation. In particular, the U-Net architecture~\cite{Ronneberger2015} has been widely used for learning-based image registration~\cite{Balakrishnan2019,XuZ2020} and further for cardiac motion estimation~\cite{ZhengQ2019,Ta2020}. 
Compared to image registration-based cardiac motion estimation, several other methods focus on anatomical motion estimation in mesh space. They explore mesh matching or mesh registration to estimate displacements for each vertex of the mesh. For example, Papademetris et al.~\cite{Papademetris2001} proposed a method that uses biomechanical modeling and shape-tracking approach to estimate the motion of myocardial mesh. Abdelkhalek et al.~\cite{Abdelkhalek2020} built a framework to compute mesh displacements via point clouds alignment. These mesh motion estimation approaches compute mesh displacements only from dynamic shape information, without considering intensity information from images. 
In contrast to the existing deep learning-based cardiac motion estimation methods, our method combines image information with the myocardial mesh which contains the epi- and endocardial surfaces of the heart. We estimate 3D mesh displacements by using the intensity information of 2D images from multiple anatomical views.

\section{Method}
Our goal is to estimate 3D mesh displacements of the LV myocardium from 2D SAX and LAX CMR images. Our task is formulated as follows: Let $\{I_0^{sa}, I_0^{2ch}, I_0^{4ch}\}$ be the 2D SAX, 2CH and 4CH view images of the ED frame and $\{I_t^{sa}, I_t^{2ch}, I_t^{4ch}\}$ be the multi-view images of the $t$-th frame. $0\leqslant t\leqslant T-1$ and $T$ is the number of frames in the cardiac cycle. We want to estimate a 3D mesh displacement $\Delta V_t\in \mathbb{R}^{N\times 3}$ for the input ED frame mesh ($\{V_0, F\}$) by using these ED and $t$-th frame multi-view images. Here, $V_0\in\mathbb{R}^{N\times 3}$ and $F$ refer to the vertices and faces of the input ED frame mesh, respectively. $N$ is the number of vertices and $\Delta V_t$ represents the motion of vertices from ED frame to the $t$-th frame.

\begin{figure}[pt]
 \centering
 \includegraphics[width=\textwidth, trim=1.6cm 11.3cm 9.5cm 0.2cm, clip]{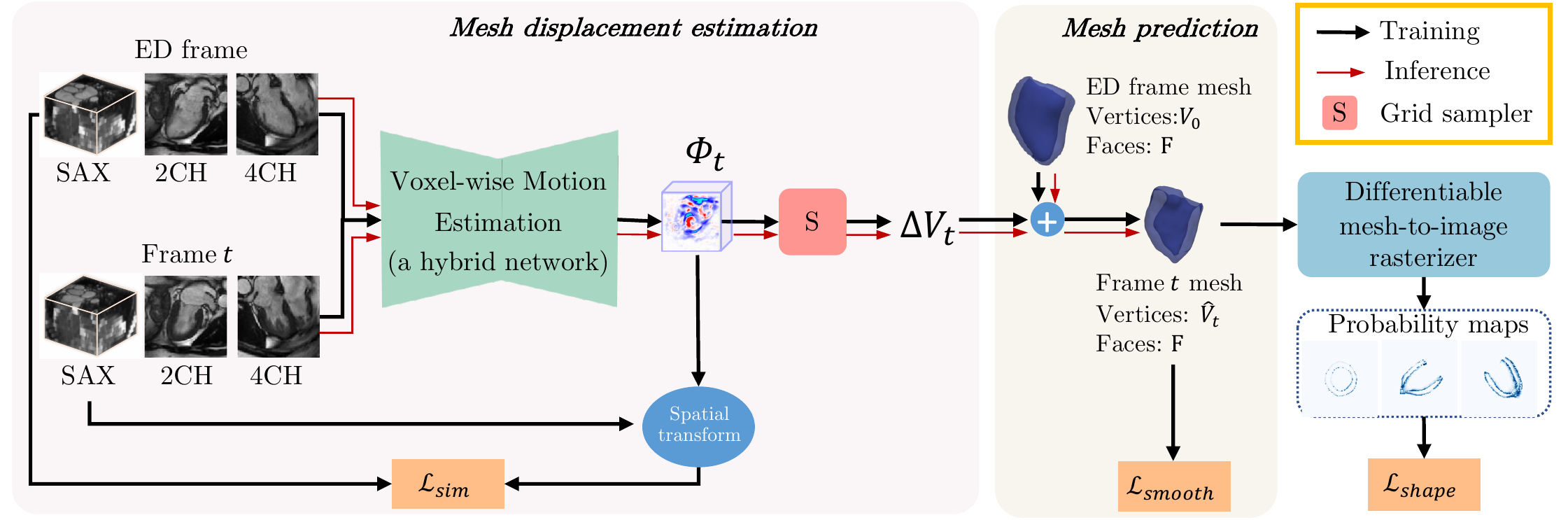}
 \caption{An overview of the proposed method. Mesh displacement estimation module takes multi-view images as input and learns mesh displacement $\Delta V_t$. By updating the myocardial mesh of the ED frame with $\Delta V_t$, mesh prediction module predicts the mesh of the $t$-th frame. During training, a differential mesh-to-image rasterizer is built to extract different anatomical view planes from the $t$-th frame mesh, which generates 2D soft segmentations (probability maps) and thus enables leveraging 2D shape information for 3D mesh displacement estimation.}
 \label{method_outline}
\end{figure}

The schematic architecture of the proposed method is shown in Fig.~\ref{method_outline}. 
For model training, the proposed method can be separated into three components: First, a mesh displacement estimation module learns the motion of a myocardial mesh from intensity images. Second, a mesh prediction module evolves the input ED frame mesh to the $t$-th frame mesh based on the learned mesh displacement. Finally, a differentiable mesh-to-image rasterizer is used to produce soft segmentations by extracting 2D planes (in the SAX and LAX orientations) from the predicted $t$-th frame mesh. This enables using 2D anatomical shape information to supervise 3D mesh displacement estimation.

\subsection{Mesh displacement estimation}
\label{displacement_estimation}
In this module, we estimate a mesh displacement $\Delta V_t$ from the input images via predicting an intermediate voxel-wise 3D motion field. In specific, a hybrid network composed of 2D CNNs and 3D CNNs learns a voxel-wise 3D motion field $\Phi_t$ from the input SAX and LAX view images. The 2D CNNs learn and combine the 2D features of multi-view images. The 3D CNNs with a encoder-decoder architecture predict $\Phi_t$ by further learning 3D representations from the combined 2D features. $\Phi_t$ has the same size to the input SAX stacks and represents the motion of image voxels from the ED frame to the $t$-th frame. Subsequently, a grid sampler is utilized to generate $\Delta V_t$ from the obtained $\Phi_t$. In detail, for each vertex of the input ED frame mesh, its displacement is sampled from $\Phi_t$ by using bi-linear interpolation at the coordinates of this vertex.

Overall, $\Delta V_t$ is estimated from the input multi-view images by
\begin{equation}\label{grid_sample}
\Delta V_t=S(H(I_0^{k}, I_{t}^{k}), V_0), \quad k=\{{{sa}}, {2ch}, {4ch}\}.
\end{equation}
Here, $H(\cdot, \cdot)$ is the hybrid network, $S(\cdot, \cdot)$ is the grid sampler and $\Phi_t=H(I_0^{k}, I_{t}^{k})$. 

As ground truth mesh displacement is usually unavailable, $\Delta V_t$ can not be directly evaluated. Instead, we evaluate $\Phi_t$ in a self supervision manner. We transform the SAX stack of the $t$-th frame ($I_t^{sa}$) to the ED frame using $\Phi_t$ via a spatial transformer~\cite{Jaderberg2015}. By minimizing the image similarity loss in Eq.~\ref{loss_sim}, $\Phi_t$ is encouraged to reflect the motion of the myocardium. 
\begin{equation}\label{loss_sim}
\mathcal{L}_{sim}=\|I_0^{sa}-I_t^{sa}\circ\Phi_t\|^2
\end{equation}

\subsection{Mesh prediction}
With the estimated $\Delta V_t$, the myocardial mesh of the ED frame ($\{V_0, F\}$) can be deformed to the $t$-th frame ($\{\hat{V}_t, F\}$) by
\begin{equation}\label{mesh_pred}
\hat{V}_t=V_0+\Delta V_t.
\end{equation}
A Laplacian smoothing loss\footnote{Implemented by pytorch3d.loss.mesh$\_$laplacian$\_$smoothing()} $\mathcal{L}_{smooth}$ is used to evaluate the smoothness of the predicted $t$-th frame mesh. The Laplacian of $\hat{v}_t^i$ is defined by $L(\hat{v}_t^i)$,
\begin{equation}\label{loss_smooth}
L(\hat{v}_t^i)=\frac{1}{|\mathcal{N}_i|}\sum_{j\in \mathcal{N}_i}(\hat{v}_t^i-\hat{v}_t^j)
\end{equation}
Here, $\mathcal{N}_i$ is the set of adjacent vertices to $\hat{v}_t^i$ and $\{\hat{v}_t^i,\hat{v}_t^j\}\in \hat{V}_t$.

\subsection{Differentiable mesh-to-image rasterizer}
Because of the low through-plane resolution in SAX stacks, the image similarity loss ($\mathcal{L}_{sim}$) on its own is not sufficient to guarantee a 3D motion estimation, especially in the longitudinal direction.
To address this problem, we propose a differentiable mesh-to-image rasterizer to extract 2D soft segmentations from the 3D mesh and thus enables 2D anatomical shape information from multiple views to supervise 3D mesh displacement estimation. These 2D soft segmentations are probability maps which contain the probability of the mesh intersecting the different view planes. The closer a vertex to a plane, the higher probability the vertex lies in the plane. The probability map enables the differentiability of the rasterization and further supports the end-to-end training of the proposed method. 
The input of the differentiable rasterizer is the mesh $\{\hat{V}_t, F\}$ and the outputs are 2D probability maps on SAX, 2CH and 4CH view planes ($\{P_t^{sa}, P_t^{2ch}, P_t^{4ch}\}$). In detail, the coordinates of $\hat{v}_t^i$ ($\hat{v}_t^i\in\hat{V}_t, i=[0,1,...,N]$) are first transformed to the image space of different anatomical planes using the information about the relative position in the DICOM header. Then, the probability of each vertex being on a specific view plane is computed by:
\begin{equation}\label{soft_slicer}
p_t^{ik}=e^{-\tau {d_{ik}^2}}, \quad d_{ik}=|z^{ik}-z^k|, \quad k=\{{{sa}}, {2ch}, {4ch}\}
\end{equation}
Here $p_t^{ik}$ refers to the probability of $\hat{v}_t^i$ belonging to the plane $k$, and  $\tau$ is the hyper-parameter which controls the sharpness the exponential function. $d_{ik}$ is the distance between $\hat{v}_t^i$ and the plane $k$, with $(x^{ik}, y^{ik}, z^{ik})$ is the transformed coordinates of $\hat{v}_t^i$ and $z^k$ is the slice corresponding to the plane $k$. The vertices satisfying $|z^{ik}-z^k|<1$ are selected as the intersection of $\{\hat{V}_t, F\}$ and plane $k$, and their probability values form the probability map $P_t^{k}$.

The obtained probability maps are compared to 2D ground truth binary segmentations $\{B_t^{sa}, B_t^{2ch}, B_t^{4ch}\}$ (only contain anatomy boundary) by a shape loss $\mathcal{L}_{shape}$. We utilize a Weighted Hausdorff Distance\footnote{Implemented by https://github.com/javiribera/locating-objects-without-bboxes}($\textrm{WHD}(\cdot,\cdot)$)~\cite{Ribera2019} to measure the distance between these anatomy boundaries,
\begin{equation}\label{loss_whd}
\mathcal{L}_{shape}=\sum\nolimits_{k=\{sa,2ch,4ch\}}\textrm{{WHD}}(P_t^k, B_t^k).
\end{equation}

\subsection{Optimization}
Our model is an end-to-end trainable framework and the overall objective is a linear combination of three loss terms, 
\begin{equation}\label{Loss}
\mathcal{L} = \mathcal{L}_{shape}+\lambda\mathcal{L}_{sim}+\beta\mathcal{L}_{smooth}, 
\end{equation}
where $\lambda$ and $\beta$ are hyper-parameters chosen experimentally depending on the dataset. We use the Adam optimizer ($\text{learning rate}=10^{-4}$) to update the network parameters. Our model is implemented by Pytorch and is trained on a NVIDIA RTX A5000 GPU with 24GB of memory.

\section{Experiments and Results}
Experiments were performed on randomly selected 530 subjects from the UK Biobank study~\cite{Petersen2015}. Each subject contains SAX, 2CH and 4CH view cine CMR sequences and each sequence contains 50 frames. SAX view images were resampled by linear interpolation from a spacing of $\sim 1.8\times 1.8\times 10mm$ to a spacing of $1.25\times 1.25\times 2mm$ while 2CH and 4CH view images were resampled from $\sim 1.8\times 1.8 mm$ to $1.25\times 1.25mm$. Based on the center of the intersecting line between the middle slice of the SAX stack and the LAX view images, the SAX, 2CH and 4CH view images are cropped to cover the whole LV in the center. The binary segmentations used in Eq.~\ref{loss_whd} were obtained via an automated  tool provided in~\cite{Duan2019}, followed by manual quality control. The myocardial meshes of the ED frame and the ES frame are reconstructed from the binary segmentations using marching cube algorithm. Each mesh contain $\sim20k$ vertices. We split the dataset into 400/50/80 for train/validation/test and train the proposed model for 300 epochs. The hyper-parameters in Eq.~\ref{Loss} are selected as $\lambda=300, \beta=200$ and $\tau$ in Eq.~\ref{soft_slicer} is set to 3. The detailed architecture of the hybrid network (Sec.~\ref{displacement_estimation}) and motion tracking videos are presented in the supplementary material. 

\begin{figure}[tb]
 \centering
 \begin{tabular}{@{\hspace{-1\tabcolsep}}c@{\hspace{0.3\tabcolsep}}c@{\hspace{0.3\tabcolsep}}c@{\hspace{0.3\tabcolsep}}c@{\hspace{0.3\tabcolsep}}c@{\hspace{0.3\tabcolsep}}c}
  \raisebox{1.5\height}{\rotatebox[origin=c]{90}{\makecell{~\scalebox{0.8}{\textbf{Mesh}}}}} &
  \includegraphics[height=1.8cm, trim=14cm 2cm 14cm 3.5cm, clip]{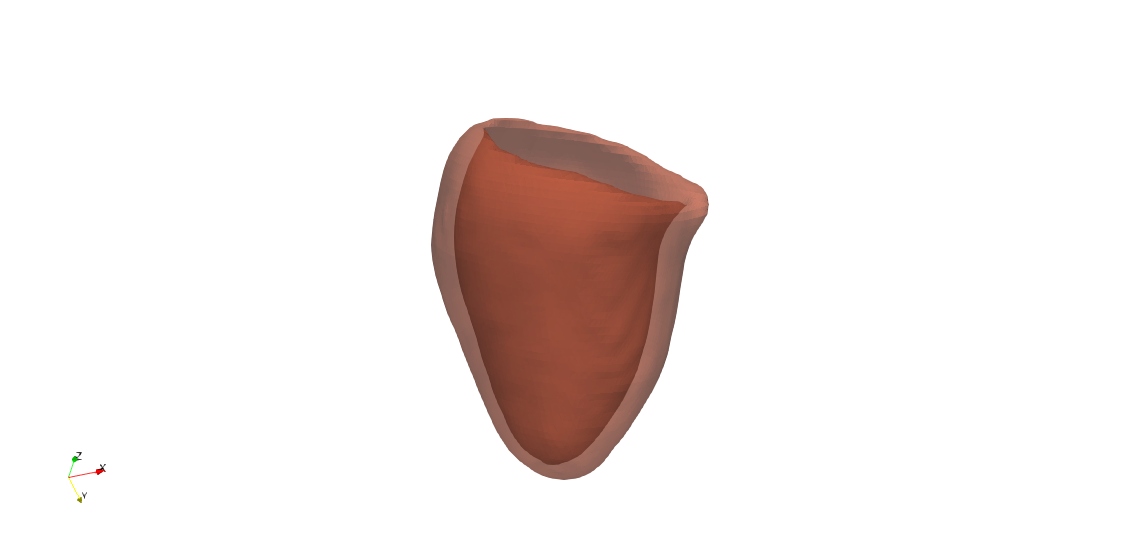} &
  \includegraphics[height=1.8cm, trim=14cm 2cm 14cm 3.5cm, clip]{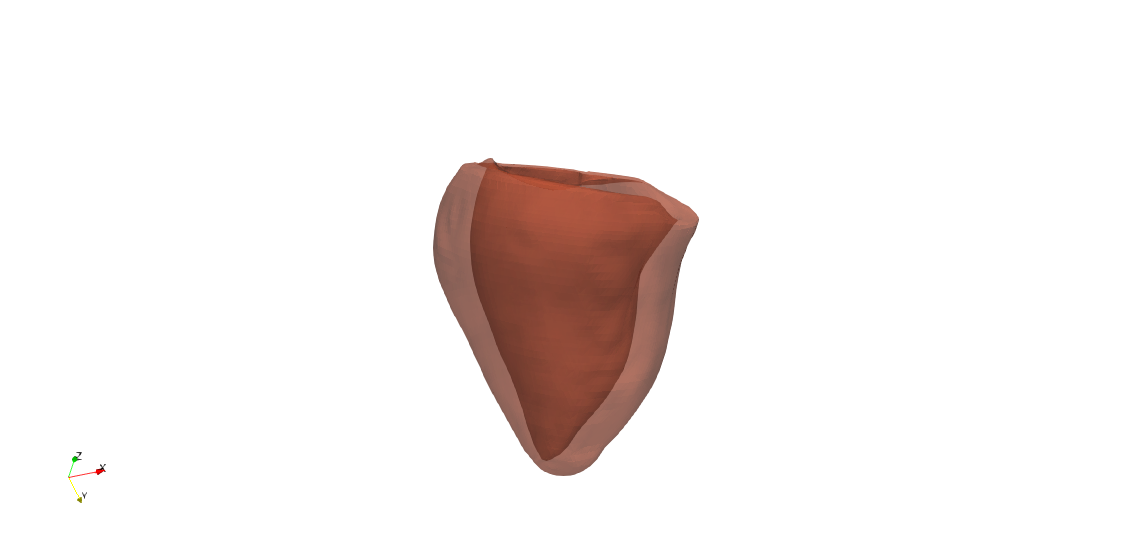} &
  \includegraphics[height=1.8cm, trim=14cm 2cm 14cm 3.5cm, clip]{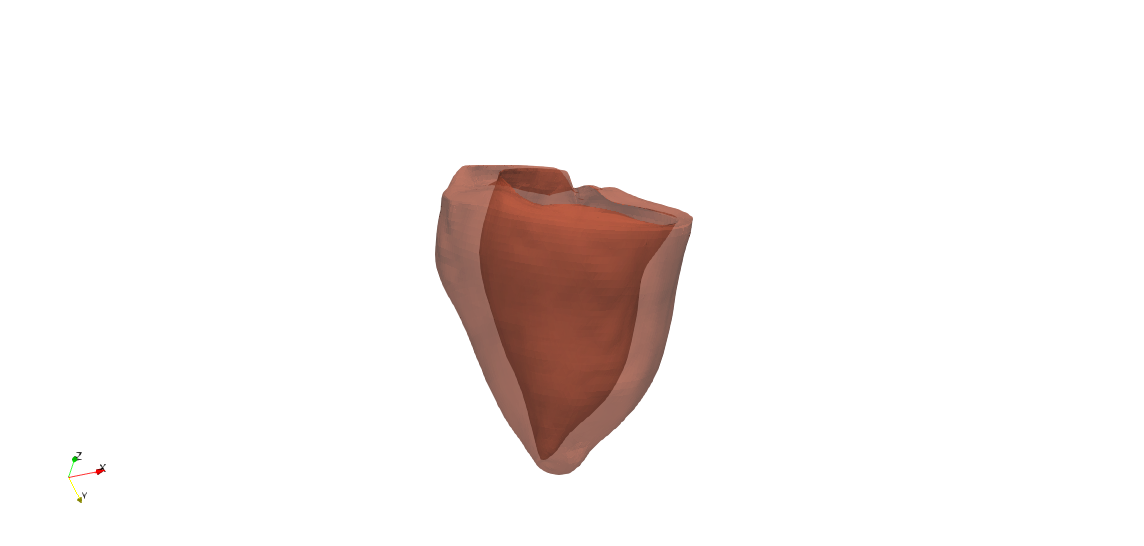} &
  \includegraphics[height=1.8cm, trim=14cm 2cm 14cm 3.5cm, clip]{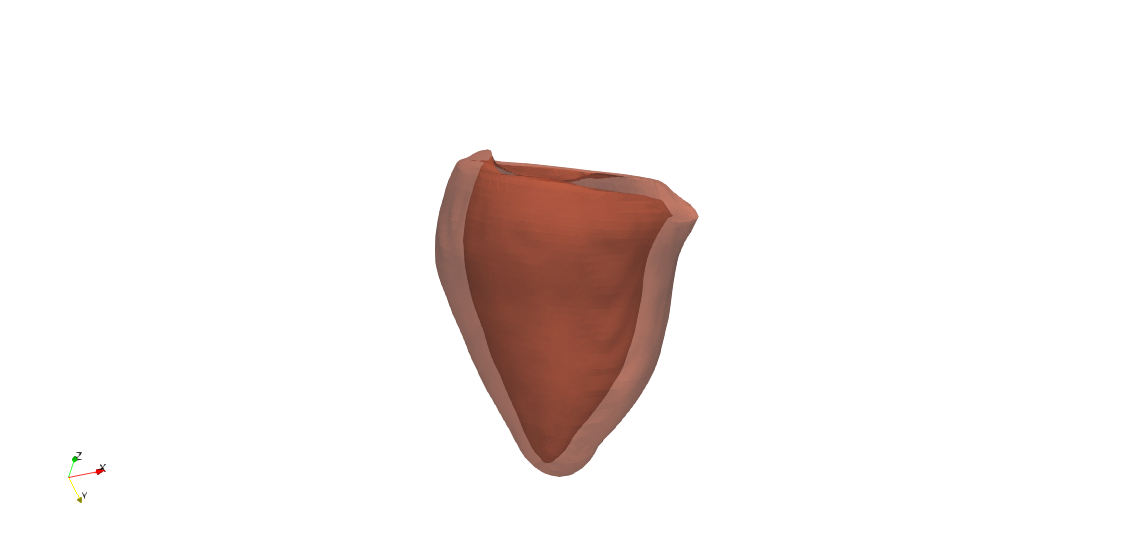} &
  \includegraphics[height=1.8cm, trim=14cm 2cm 14cm 3.5cm, clip]{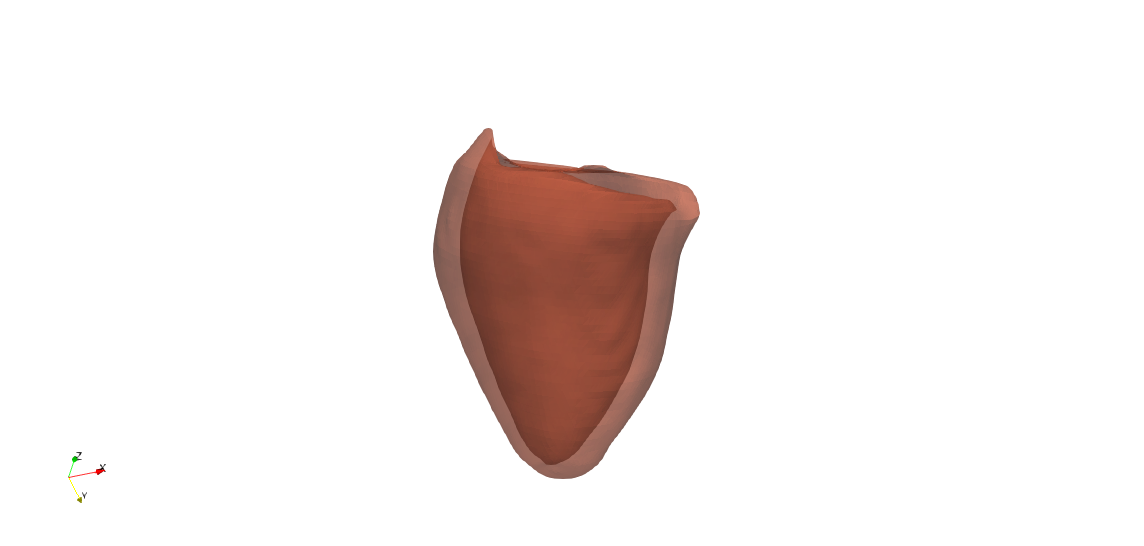} \\
  \raisebox{2\height}{\rotatebox[origin=c]{90}{\makecell{~\scalebox{0.8}{\textbf{SAX}}}}} &
  \includegraphics[height=1.9cm]{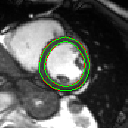} &
  \includegraphics[height=1.9cm]{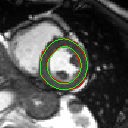} &
  \includegraphics[height=1.9cm]{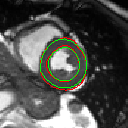} &
  \includegraphics[height=1.9cm]{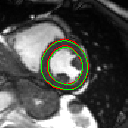} &
  \includegraphics[height=1.9cm]{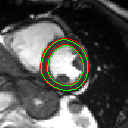} \\
  \raisebox{2\height}{\rotatebox[origin=c]{90}{\makecell{~\scalebox{0.8}{\textbf{2CH}}}}} &
  \includegraphics[height=1.9cm]{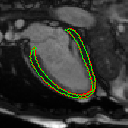} &
  \includegraphics[height=1.9cm]{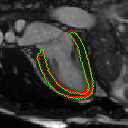} &
  \includegraphics[height=1.9cm]{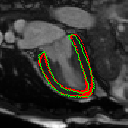} &
  \includegraphics[height=1.9cm]{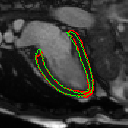} &
  \includegraphics[height=1.9cm]{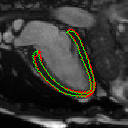} \\
  \raisebox{2\height}{\rotatebox[origin=c]{90}{\makecell{~\scalebox{0.8}{\textbf{4CH}}}}} &
  \includegraphics[height=1.9cm]{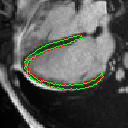} &
  \includegraphics[height=1.9cm]{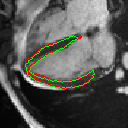} &
  \includegraphics[height=1.9cm]{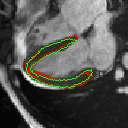} &
  \includegraphics[height=1.9cm]{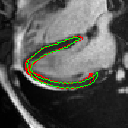} &
  \includegraphics[height=1.9cm]{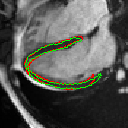} \\
  ~~~ &
  \raisebox{0.1\height}{\rotatebox[origin=c]{0}{\makecell{~\scalebox{0.8}{\textbf{t=0}}}}} &
  \raisebox{0.1\height}{\rotatebox[origin=c]{0}{\makecell{~\scalebox{0.8}{\textbf{t=10}}}}} &
  \raisebox{0.1\height}{\rotatebox[origin=c]{0}{\makecell{~\scalebox{0.8}{\textbf{t=20}}}}} &
  \raisebox{0.1\height}{\rotatebox[origin=c]{0}{\makecell{~\scalebox{0.8}{\textbf{t=30}}}}} &
  \raisebox{0.1\height}{\rotatebox[origin=c]{0}{\makecell{~\scalebox{0.8}{\textbf{t=40}}}}}
  \end{tabular}
  \caption{Examples of motion tracking results. The ED frame mesh is deformed to the $t$-th frame using the mesh displacements generated by the proposed method. 2D segmentation on SAX, 2CH and 4CH view planes (Row 2-4) are generated by extracting the corresponding planes from the predicted $t$-th frame mesh. Red contours are predicted segmentation while green contours are ground truth.}
  \label{singlesub_results}
\end{figure}

\textit{\textbf{Mesh motion tracking performance.}}
The proposed method is utilized to estimate mesh displacements in the full cardiac cycle. For each test subject, with the obtained $\{\Delta V_t|t=[0,49]\}$, the myocardial mesh of the ED frame ($t=0$) is deformed to the $t$-th frame. Red meshes in Fig.~\ref{singlesub_results} shows that the estimated mesh displacement $\Delta V_t$ enables 3D myocardial motion tracking on meshes. In addition, we extracted SAX/2CH/4CH view planes from the predicted $t$-th frame mesh and generated the predicted 2D segmentations on different view planes. Fig.~\ref{singlesub_results} shows the effectiveness of $\Delta V_t$ by comparing the predicted and the ground truth 2D myocardial segmentations (only contain anatomy boundaries).

\begin{table}[tb]
\centering
\caption{Comparison of other cardiac motion tracking methods. The results are reported as ``mean (standard deviation)". $\uparrow$ indicates the higher value the better while $\downarrow$ indicates the lower value the better. Best results in bold.}
\label{quantitative_comparison}
\resizebox{\textwidth}{!}{
\begin{tabular}{lcc|ccc|ccc}
\toprule[1.2pt]
\multirow{2}{*}{Methods}                        & 
\multirow{2}{*}{\makecell{Anatomical\\ views}}               &
\multirow{2}{*}{\makecell{Mean Surface \\ distance} $\downarrow$}               &
\multicolumn{3}{c|}{HD (mm) $\downarrow$ }            &
\multicolumn{3}{c}{BoundF ($\%$) $\uparrow$ }            \\
\cmidrule{4-9}
~~~~~ &
~~~~~ &
~~~~~ &
SAX &
2CH &
4CH &
SAX &
2CH &
4CH \\
\midrule
FFD~\cite{Rueckert1999}                        &
SAX                                              &
3.02(0.86) &
21.31(4.66) &
15.17(4.52) &
15.95(4.84) &
62.15(7.48) &
77.60(6.97) &
77.79(7.13) \\
dDemons~\cite{Vercauteren2007}                            &
SAX                                  &
3.20(0.90) &
\textbf{20.32(5.03)} &
15.01(3.48) &
15.72(3.41) &
63.67(6.92) &
77.38(5.99) &
80.29(5.83) \\
3D-UNet~\cite{ociek2016}                                &
SAX                                             &
3.35(0.88) &
20.45(5.15) &
14.44(2.99) &
14.83(3.57) &
60.64(7.74) &
74.63(6.01) &
76.06(6.08) \\
Ours                                           &
SAX, 2CH, 4CH                                              &
\textbf{1.98(0.44)} &
20.76(4.82) &
\textbf{7.44(4.04)} &
\textbf{8.62(4.49)} &
\textbf{71.49(8.82)} &
\textbf{87.21(6.97)}&
\textbf{84.24(6.84)} \\
\bottomrule[1.2pt]
\end{tabular}
}
\end{table}

\begin{figure}[ptb]
 \centering
 \begin{tabular}{@{\hspace{-1\tabcolsep}}c@{\hspace{0.5\tabcolsep}}c@{\hspace{0.5\tabcolsep}}c@{\hspace{0.5\tabcolsep}}c@{\hspace{0.5\tabcolsep}}c@{\hspace{0.5\tabcolsep}}c}
  \includegraphics[height=2cm, trim=14cm 2cm 14cm 3.5cm, clip]{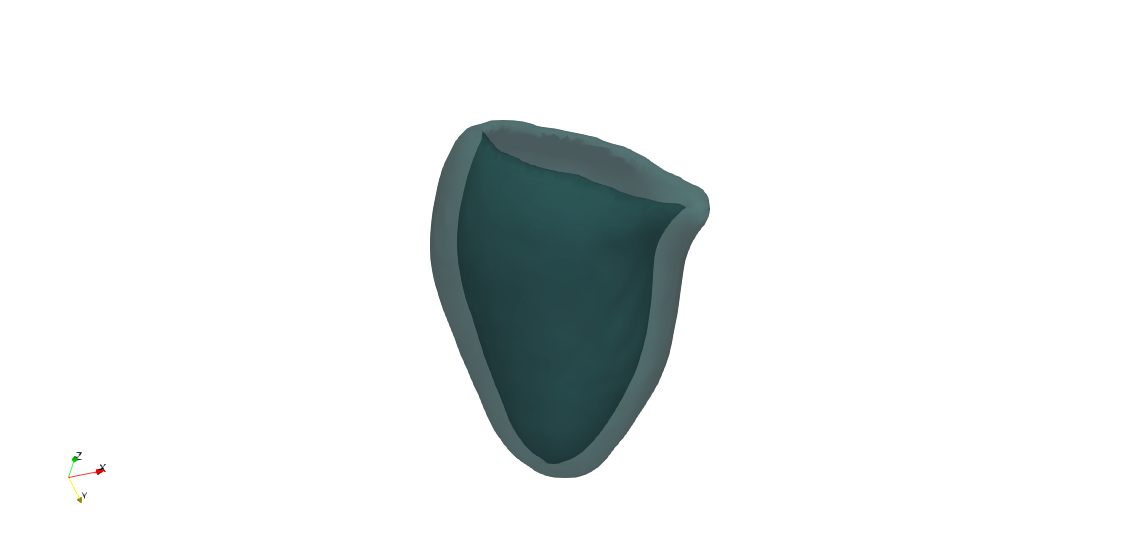} &
  \includegraphics[height=2cm, trim=14cm 2cm 14cm 3.5cm, clip]{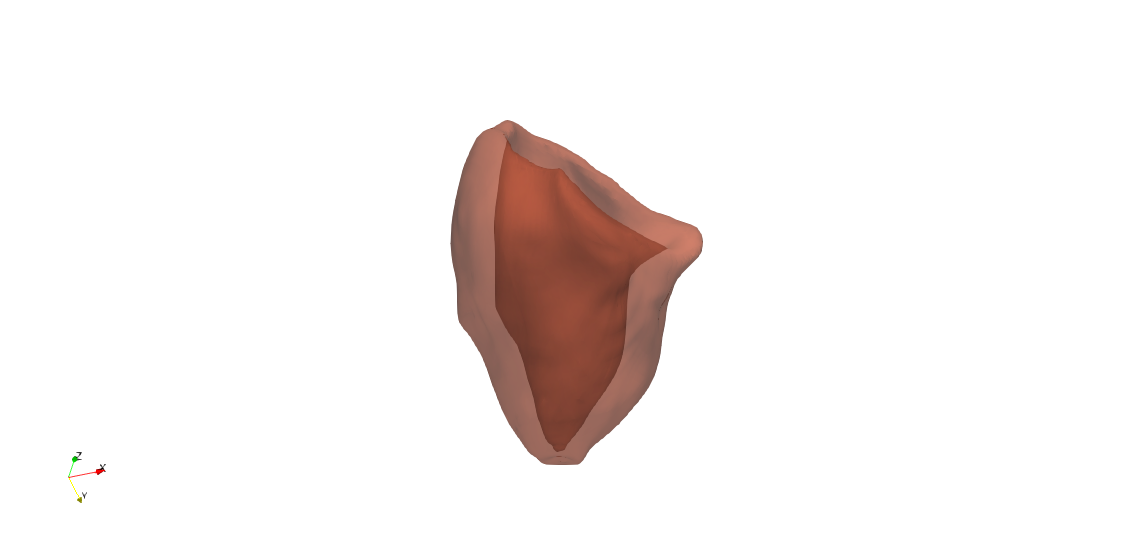} &
  \includegraphics[height=2cm, trim=14cm 2cm 14cm 3.5cm, clip]{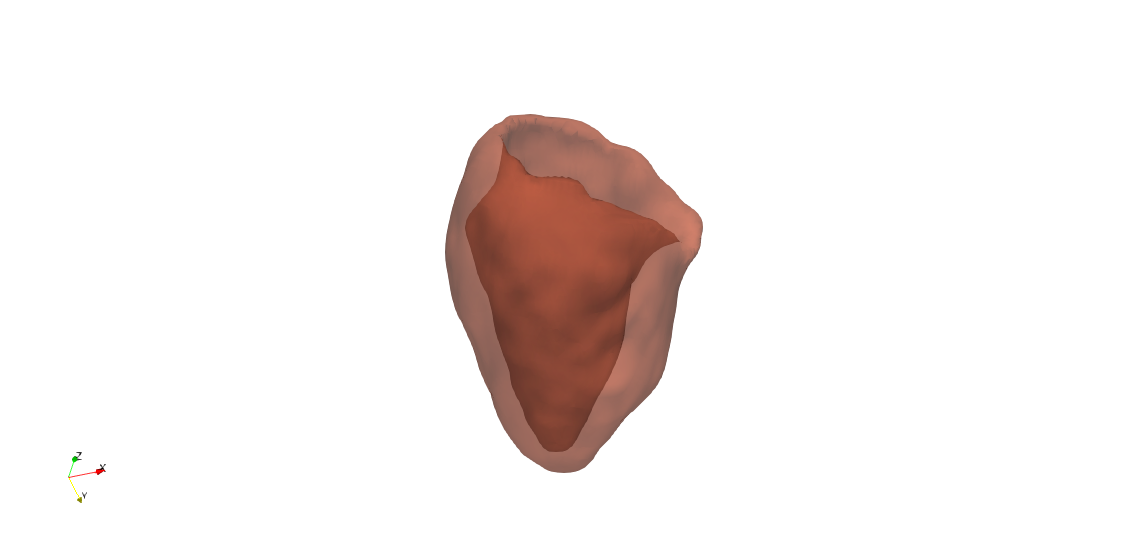} &
  \includegraphics[height=2cm, trim=14cm 2cm 14cm 3.5cm, clip]{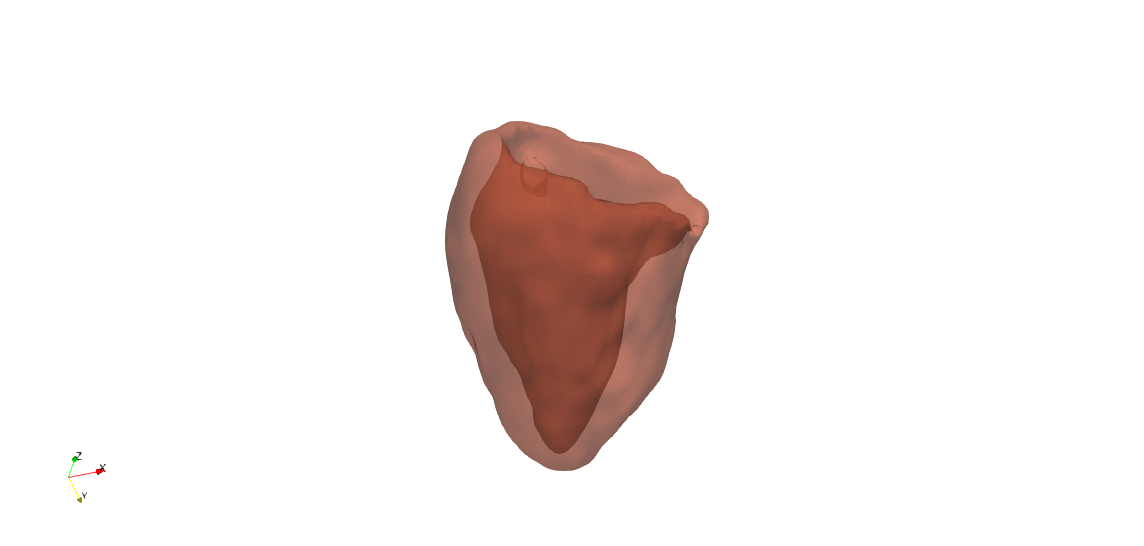} &
  \includegraphics[height=2cm, trim=14cm 2cm 14cm 3.5cm, clip]{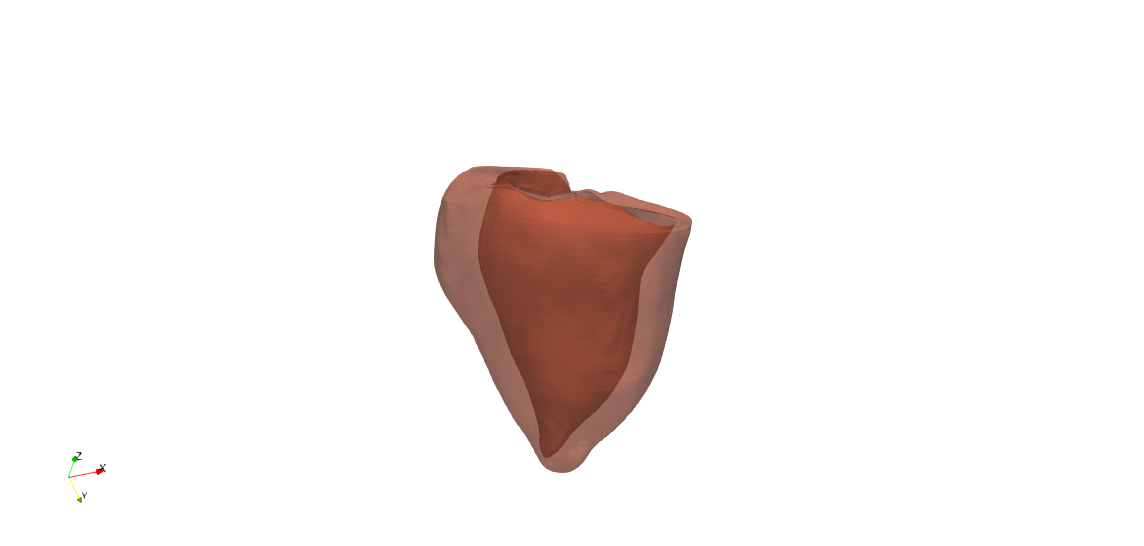} &
  \includegraphics[height=2cm, trim=14cm 2cm 14cm 3.5cm, clip]{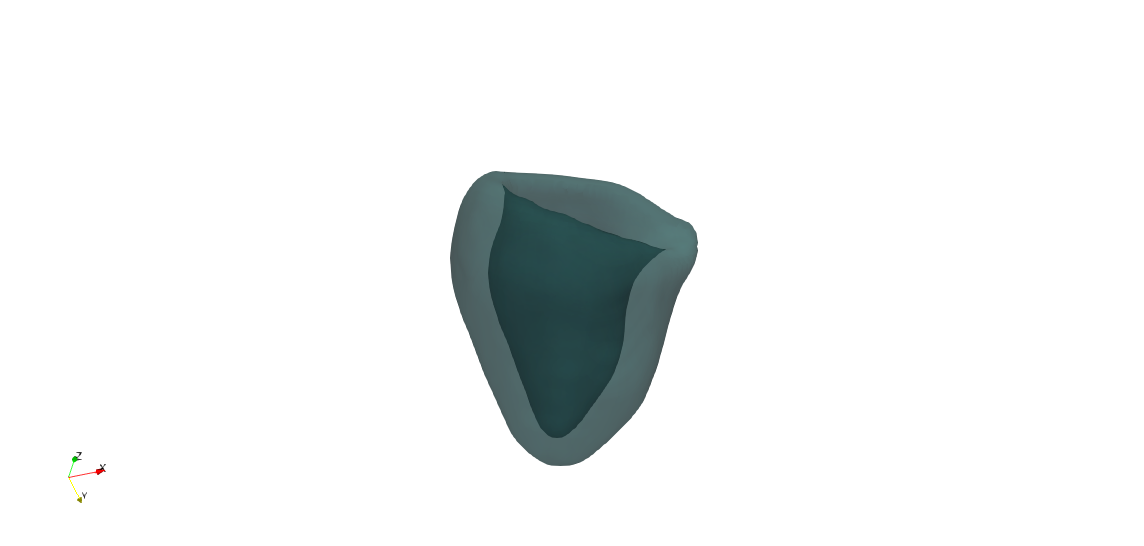} \\
  
  \raisebox{0.1\height}{\rotatebox[origin=c]{0}{\makecell{~\scalebox{0.8}{\textbf{ED frame (GT)}}}}} &
  \raisebox{0.1\height}{\rotatebox[origin=c]{0}{\makecell{~\scalebox{0.8}{\textbf{FFD~\cite{Rueckert1999}}}}}} &
  \raisebox{0.1\height}{\rotatebox[origin=c]{0}{\makecell{~\scalebox{0.8}{\textbf{dDemons~\cite{Vercauteren2007}}}}}} &
  \raisebox{0.1\height}{\rotatebox[origin=c]{0}{\makecell{~\scalebox{0.8}{\textbf{3D-UNet~\cite{ociek2016}}}}}} &
  \raisebox{0.1\height}{\rotatebox[origin=c]{0}{\makecell{~\scalebox{0.8}{\textbf{Ours}}}}} &
  \raisebox{0.1\height}{\rotatebox[origin=c]{0}{\makecell{~\scalebox{0.8}{\textbf{ES frame (GT)}}}}}
  \end{tabular}
  \caption{Motion estimation using baseline methods and the proposed method. Green meshes are ground truth (GT) meshes of the ED and ES frames. Red meshes are the predicted ES frame meshes based on different methods.}
  \label{mesh_comparison}
\end{figure}

\textit{\textbf{Comparison study.}}
We compared the proposed method with three state-of-the-art cardiac motion tracking approaches, including a B-spline free form deformation (FFD) algorithm\footnote{Implemented by using the MIRTK toolkit: http://mirtk.github.io/}~\cite{Rueckert1999},  a diffeomorphic Demons (dDemons) algorithm\footnote{ https://github.com/InsightSoftwareConsortium/SimpleITK-Notebooks/Python/66}~\cite{Vercauteren2007} and 3D-UNet\footnote{https://github.com/wolny/pytorch-3dunet}~\cite{ociek2016}. 
We quantitatively evaluated the performance using the following metrics: Mean surface distance, Hausdorff distance (HD) and Boundary F-score (BoundF). 
Here, surface distance evaluates the distance between the predicted and the ground truth ES frame meshes. The Hausdorff distance and Boundary F-score compare the predicted and the ground truth 2D myocardial segmentations on SAX, 2CH and 4CH view planes. The Hausdorff distance quantifies the contour distance while Boundary F-score evaluates contour alignment accuracy as described in~\cite{Perazzi2016,Cheng2019,Gur2020}.
For fair comparison, we evaluated several sets of hyper-parameter values for all methods and selected hyper-parameters that achieve the best Hausdorff distance on the validation set. 
Table~\ref{quantitative_comparison} shows the quantitative comparison results and Fig.~\ref{mesh_comparison} further shows the qualitatively results. From Table~\ref{quantitative_comparison}, we observe that the proposed method outperforms all baseline methods, demonstrating the effectiveness of the proposed method for estimating mesh displacements. In addition, the proposed method achieves the best performance regarding to 2CH and 4CH view segmentations in Table~\ref{quantitative_comparison} and obtains the ES frame mesh which is most similar to the ground truth ES frame mesh in Fig.~\ref{mesh_comparison}. These results illustrate that in contrast to all baseline methods which only show motion within SAX plane, the proposed method is able to estimate through-plane motion along the longitudinal direction. Finally, we compared the average inference time of different methods. The FFD~\cite{Rueckert1999} is $16s$ and the dDemons~\cite{Vercauteren2007} is $28s$ (both are only available on CPUs) while the 3D-UNet~\cite{ociek2016} is $1.1s$ and our method is $1.0s$ (both are on GPUs).

\textit{\textbf{Ablation study.}}
For the proposed method, we explore the effects of using different anatomical views and loss combinations. Table~\ref{diff_views} shows that adding the images and shape information of LAX views improves the performance. This might be because LAX views can introduce high-resolution through-plane information for 3D motion estimation. Table~\ref{diff_loss} shows that proposed method performs best, which illustrates the importance of each loss component.

\begin{table}[t]
  \begin{minipage}[t]{0.49\textwidth} 
    \caption{Mesh displacement estimation with \textbf{different anatomical views}.}
    \label{diff_views}
    \resizebox{\textwidth}{!}{
    \begin{tabular}{lccc}
    \toprule[1.2pt]
    \multirow{2}{*}{Anatomical views}                 &
    \multicolumn{3}{c}{HD (mm) $\downarrow$ }            \\
    \cmidrule{2-4}
    ~~~~~ &
    SAX &
    2CH &
    4CH \\
    \midrule
    SAX                                              &
    21.09(4.56) &
    9.29(4.80) &
    12.12(5.51) \\
    SAX$+$2CH                                &
    \textbf{20.35(5.41)} &
    9.99(4.36) &
    11.98(4.14) \\
    SAX$+$2CH$+$4CH    &
    20.76(4.82) &
    \textbf{7.44(4.04)} &
    \textbf{8.62(4.49)} \\
    \bottomrule[1.2pt]
    \end{tabular}
    }
  \end{minipage}%
  \hfill
  \begin{minipage}[t]{0.49\textwidth}
    \caption{Mesh displacement estimation with \textbf{different losses}.}
    \label{diff_loss}
    \resizebox{\textwidth}{!}{
    \begin{tabular}{lccc}
    \toprule[1.2pt]
    \multirow{2}{*}{Loss}                 &
    \multicolumn{3}{c}{HD (mm) $\downarrow$ }            \\
    \cmidrule{2-4}
    ~~~~~ &
    SAX &
    2CH &
    4CH \\
    \midrule
    $\mathcal{L}_{shape}$ &                             
    \textbf{20.21(5.12)} &
    13.31(2.77) &
    14.36(4.24) \\
    $\mathcal{L}_{shape}+\mathcal{L}_{smooth}$ &    
    20.73(4.73) &
    12.19(3.09) &
    14.05(3.26) \\
    $\mathcal{L}_{shape}+\mathcal{L}_{smooth}+\mathcal{L}_{sim}$    &
    20.76(4.82) &
    \textbf{7.44(4.04)} &
    \textbf{8.62(4.49)} \\
    \bottomrule[1.2pt]
    \end{tabular}
    }
  \end{minipage}
\end{table}

\textit{\textbf{Discussion.}}
An alternative to our method would be to estimate mesh displacement directly from input images via fully connected networks (FCNs) without voxel-wise 3D motion estimation. However, using FCNs to estimate displacement of $\sim 20K$ vertices needs large GPU memory, which may not always be available. More importantly, voxel-wise 3D motion estimation is able to explicitly connect the anatomical motion in image space to mesh space. 
$\mathcal{L}_{smooth}$ is introduced as an overall regularization term to directly constrain the smoothness of the deformed mesh. The regularization loss for the voxel-wise 3D motion field could be added. But the extra loss may increase the complexity of model training.
In contrast to the commonly used 3D-to-2D projection which obtains a silhouette by utilizing camera parameters, \emph{e.g.},~\cite{Tung2017,Kato2018}, our differentiable mesh-to-image rasterizer focuses on slicing a 3D mesh to 2D soft segmentations.
Weighted Hausdorff Distance is one example to compute $\mathcal{L}_{shape}$, as it can evaluate the distance between soft-labeled and hard-labeled point sets. Other boundary similarity measurements may also be applied to this loss component in our task. 
In the evaluation, we quantitatively evaluated the performance on the ES frame. This is because ground truth meshes are only available at the ED and ES frames in our current dataset.
We choose hyper-parameters using grid search (\emph{i.e.}, $\lambda=[100,300,500]$, $\beta=[100,200,300]$ and $\tau=[2,3]$) and we select the hyper-parameters with the best performance on the validation data.

\section{Conclusion}
In this paper, we propose a image-based mesh motion estimation network for 3D myocardial motion tracking. The proposed method is an end-to-end trainable network which estimates 3D mesh displacements from the intensity information of 2D SAX and LAX view CMR images. Experimental results demonstrate the effectiveness of the proposed method compared with other competing methods. For future work, we will apply our method to clinical association studies.

\subsubsection*{Acknowledgments.} This research has been conducted using the UK Biobank Resource under application number 40616. This work is supported by the British Heart Foundation (RG/19/6/34387, RE/18/4/34215); Medical Research Council (MC-A658-5QEB0); National Institute for Health Research (NIHR) Imperial College Biomedical Research Centre. W. Bai was supported by EPSRC DeepGeM Grant (EP/W01842X/1).


%
%
%
\bibliographystyle{splncs04}

\begin{thebibliography}{10}
\providecommand{\url}[1]{\texttt{#1}}
\providecommand{\urlprefix}{URL }
\providecommand{\doi}[1]{https://doi.org/#1}

\bibitem{Abdelkhalek2020}
Abdelkhalek, M., Aguib, H., Moustafa, M., Elkhodary, K.: Enhanced {3D}
  myocardial strain estimation from multi-view 2{D} {CMR} imaging. arXiv
  preprint arXiv:2009.12466  (2020)

\bibitem{Bai2020}
Bai, W., Suzuki, H., Huang, J., Francis, C., Wang, S., Tarroni, G., Guitton,
  F., Aung, N., Fung, K., Petersen, S., Piechnik, S., Neubauer, S., Evangelou,
  E., Dehghan, A., O'Regan, D., Wilkins, M., Guo, Y., Matthews, P., Rueckert,
  D.: A population-based phenome-wide association study of cardiac and aortic
  structure and function. Nat Med  \textbf{26},  1654--1662 (2020)

\bibitem{Balakrishnan2019}
Balakrishnan, G., Zhao, A., Sabuncu, M.R., Guttag, J.V., Dalca, A.V.:
  Voxelmorph: {A} learning framework for deformable medical image registration.
  IEEE Trans Med Imaging  \textbf{38}(8),  1788--1800 (2019)

\bibitem{Bello2019}
Bello, G., Dawes, T., Duan, J., Biffi, C., de~Marvao, A., Howard, L., Gibbs,
  S., Wilkins, M., Cook, S., Rueckert, D., O'Regan, D.P.: Deep learning cardiac
  motion analysis for human survival prediction. Nat Mach Intell  \textbf{1},
  95--104 (2019)

\bibitem{Cheng2019}
Cheng, D., Liao, R., Fidler, S., Urtasun, R.: Darnet: Deep active ray network
  for building segmentation. In: CVPR. pp. 7423--7431 (2019)

\bibitem{ociek2016}
{\c{C}}i{\c{c}}ek, {\"O}., Abdulkadir, A., Lienkamp, S., Brox, T., Ronneberger,
  O.: 3{D} {U}-net: Learning dense volumetric segmentation from sparse
  annotation. In: MICCAI. pp. 424--432 (2016)

\bibitem{Claus2015}
Claus, P., Omar, A.M.S., Pedrizzetti, G., Sengupta, P.P., Nagel, E.: Tissue
  tracking technology for assessing cardiac mechanics: Principles, normal
  values, and clinical applications. JACC Cardiovasc Imaging  \textbf{8}(12),
  1444--1460 (2015)

\bibitem{Craene2012}
Craene, M.D., Piella, G., Camara, O., Duchateau, N., Silva, E., Doltra, A.,
  D’hooge, J., Brugada, J., Sitges, M., Frangi, A.F.: Temporal diffeomorphic
  free-form deformation: Application to motion and strain estimation from 3{D}
  echocardiography. Med Imag Anal  \textbf{16}(2),  427--450 (2012)

\bibitem{Duan2019}
Duan, J., Bello, G., Schlemper, J., Bai, W., Dawes, T.J.W., Biffi, C.,
  de~Marvao, A., Doumoud, G., O’Regan, D.P., Rueckert, D.: Automatic 3{D}
  bi-ventricular segmentation of cardiac images by a shape-refined multi- task
  deep learning approach. IEEE Trans Med Imaging  \textbf{38}(9),  2151--2164
  (2019)

\bibitem{Gur2020}
Gur, S., Shaharabany, T., Wolf, L.: End to end trainable active contours via
  differentiable rendering. In: ICLR (2020)

\bibitem{Jaderberg2015}
Jaderberg, M., Simonyan, K., Zisserman, A., Kavukcuoglu, K.: Spatial
  transformer networks. In: NeurIPS (2015)

\bibitem{Kato2018}
Kato, H., Ushiku, Y., Harada, T.: Neural 3d mesh renderer. In: CVPR (2018)

\bibitem{Loecher2021}
Loecher, M., Perotti, L.E., Ennis, D.B.: Using synthetic data generation to
  train a cardiac motion tag tracking neural network. Med Imag Anal
  \textbf{74} (2021)

\bibitem{Mansi2011}
Mansi, T., Voigt, I., Leonardi, B., Pennec, X., Durrleman, S., Sermesant, M.,
  Delingette, H., Taylor, A.M., Boudjemline, Y., Pongiglione, G., Ayache, N.: A
  statistical model for quantification and prediction of cardiac remodelling:
  Application to {T}etralogy of {F}allot. IEEE Trans Med Imaging
  \textbf{30}(9),  1605--1616 (2011)

\bibitem{Marvao2015}
de~Marvao, A., Dawes, T.J.W., Shi, W., Durighel, G., Rueckert, D., Cook, S.A.,
  O’Regan, D.P.: Precursors of the hypertensive heart phenotype develop in
  normotensive adults: a high resolution 3{D} {MRI} study. JACC Cardiovasc
  Imaging  (2015)

\bibitem{McLeod2011}
McLeod, K., Prakosa, A., Mansi, T., Sermesant, M., Pennec, X.: An
  incompressible log-domain demons algorithm for tracking heart tissue. In:
  MICCAI workshop STACOM (2011)

\bibitem{Papademetris2001}
Papademetris, X., Sinusas, A.J., Dione, D.P., Duncan, J.S.: Estimation of 3{D}
  left ventricular deformation from echocardiography. Med Imag Anal
  \textbf{5}(1),  17--28 (2001)

\bibitem{Perazzi2016}
Perazzi, F., Pont-Tuset, J., McWilliams, B., Van~Gool, L., Gross, M.,
  Sorkine-Hornung, A.: A benchmark dataset and evaluation methodology for video
  object segmentation. In: CVPR (2016)

\bibitem{Petersen2015}
Petersen, S., Matthews, P., Francis, J., Robson, M., Zemrak, F., Boubertakh,
  R., Young, A., Hudson, S., Weale, P., Garratt, S., Collins, R., Piechnik, S.,
  Neubauer, S.: {UK B}iobank’s cardiovascular magnetic resonance protocol. J
  Cardiovasc Magn Reson  \textbf{18} (2015)

\bibitem{Puyol2019}
Puyol-Antón, E., Ruijsink, B., Gerber, B., Amzulescu, M.S., Langet, H.,
  De~Craene, M., Schnabel, J.A., Piro, P., King, A.P.: Regional multi-view
  learning for cardiac motion analysis: Application to identification of
  dilated cardiomyopathy patients. IEEE Trans Biomed Eng  \textbf{66}(4),
  956--966 (2019)

\bibitem{Qin2018}
Qin, C., Bai, W., Schlemper, J., Petersen, S.E., Piechnik, S.K., Neubauer, S.,
  Rueckert, D.: Joint learning of motion estimation and segmentation for
  cardiac {MR} image sequences. In: MICCAI (2018)

\bibitem{Qin2020}
Qin, C., Wang, S., Chen, C., Qiu, H., Bai, W., Rueckert, D.:
  Biomechanics-informed neural networks for myocardial motion tracking in
  {MRI}. In: MICCAI (2020)

\bibitem{Ribera2019}
Ribera, J., Guera, D., Chen, Y., Delp, E.J.: Locating objects without bounding
  boxes. In: CVPR. pp. 6472--6482 (2019)

\bibitem{Ronneberger2015}
Ronneberger, O., P.Fischer, Brox, T.: U-{N}et: {C}onvolutional networks for
  biomedical image segmentation. In: MICCAI. pp. 234--241 (2015)

\bibitem{Rueckert1999}
Rueckert, D., Sonoda, L., Hayes, C., Hill, D., Leach, M., Hawkes, D.: Nonrigid
  registration using free-form deformations: application to breast {MR} images.
  IEEE Trans Med Imaging  \textbf{18}(8),  712--721 (1999)

\bibitem{Shi2012}
Shi, W., Zhuang, X., Wang, H., Duckett, S., Luong, D.V.N., Tobon-Gomez, C.,
  Tung, K.P., Edwards, P., Rhode, K., Razavi, R., Ourselin, S., Rueckert, D.: A
  comprehensive cardiac motion estimation framework using both untagged and
  3-{D} tagged {MR} images based on nonrigid registration. IEEE Trans Med
  Imaging  \textbf{31},  1263--1275 (2012)

\bibitem{Ta2020}
Ta, K., Ahn, S.S., Stendahl, J.C., Sinusas, A.J., Duncan, J.S.: A
  semi-supervised joint network for simultaneous left ventricular motion
  tracking and segmentation in 4{D} echocardiography. In: MICCAI (2020)

\bibitem{Tobon2013}
Tobon-Gomez, C., {De Craene}, M., McLeod, K., Tautz, L., Shi, W., Hennemuth,
  A., Prakosa, A., Wang, H., Carr-White, G., Kapetanakis, S., Lutz, A., Rasche,
  V., Schaeffter, T., Butakoff, C., Friman, O., Mansi, T., Sermesant, M.,
  Zhuang, X., Ourselin, S., Peitgen, H.O., Pennec, X., Razavi, R., Rueckert,
  D., Frangi, A., Rhode, K.: Benchmarking framework for myocardial tracking and
  deformation algorithms: An open access database. Med Imag Anal
  \textbf{17}(6),  632--648 (2013)

\bibitem{Tung2017}
Tung, H.Y.F., Tung, H.W., Yumer, E., Fragkiadaki, K.: Self-supervised learning
  of motion capture. In: NeurIPS (2017)

\bibitem{Vercauteren2007}
Vercauteren, T., Pennec, X., Perchant, A., Ayache, N.: Non-parametric
  diffeomorphic image registration with the demons algorithm. In: MICCAI (2007)

\bibitem{XuZ2020}
Xu, Z., Luo, J., Yan, J., Pulya, R., Li, X., Wells, W., Jagadeesan, J.:
  Adversarial uni- and multi-modal stream networks for multimodal image
  registration. In: MICCAI. pp. 222--232 (2020)

\bibitem{Ye2021}
Ye, M., Kanski, M., Yang, D., Chang, Q., Yan, Z., Huang, Q., Axel, L., Metaxas,
  D.: Deeptag: An unsupervised deep learning method for motion tracking on
  cardiac tagging magnetic resonance images. In: CVPR (2021)

\bibitem{Yu2020}
Yu, H., Sun, S., Yu, H., Chen, X., Shi, H., Huang, T.S., Chen, T.: {FOAL}: Fast
  online adaptive learning for cardiac motion estimation. In: CVPR (2020)

\bibitem{ZhengQ2019}
Zheng, Q., Delingette, H., Ayache, N.: Explainable cardiac pathology
  classification on cine {MRI} with motion characterization by semi-supervised
  learning of apparent flow. Med Imag Anal  \textbf{56},  80--95 (2019)

\end{thebibliography}

%
\end{document}